\documentclass[11pt]{article}
\usepackage[T1]{fontenc}
\usepackage[utf8]{inputenc}
\usepackage{geometry}
\usepackage{float}
\usepackage{textcomp}
\usepackage{amsmath}
\usepackage{graphicx}
\usepackage{subfig}
\usepackage{setspace}
\usepackage{color}
\usepackage{titlesec}

\makeatletter

\newcommand{\lyxmathsym}[1]{\ifmmode\begingroup\def\b@ld{bold}
	\text{\ifx\math@version\b@ld\bfseries\fi#1}\endgroup\else#1\fi}

\providecommand{\tabularnewline}{\\}

\usepackage{slashed}
\usepackage[colorlinks=true]{hyperref}

\usepackage[numbers,sort&compress]{natbib}

\makeatother

\usepackage{babel}

\begin{document}
	\begin{sloppypar}

\begin{titlepage}
\title{\huge{\bf{$\Lambda_{b}\rightarrow P\ell$ factorization in QCD}}}
	\date{}
\maketitle
\vspace{-12mm}
\thispagestyle{empty}
\begin{center}       
\author{ \bf{Lei-Yi Li$\,^{a,b}\,$\footnote{Corresponding author: lily@ihep.ac.cn},
Cai-Dian L\"u$\,^{a,b}\,$\footnote{Corresponding author: lucd@ihep.edu.cn}, 
Jin Wang$\,^{a,b}\,$\footnote{Corresponding author: wangjin99@ihep.ac.cn},
Yan-Bing Wei$\,^{c}\,$\footnote{Corresponding author: yanbing.wei@bjut.edu.cn}
}}\\
  \vspace{2mm}
{ \it $^a$ Institute of High Energy Physics, CAS, P.O. Box 918(4) Beijing 100049,  China} \\         
{ \it $^b$ School of Physical Sciences, University of Chinese Academy of Sciences, Beijing 100049, China} \\         
{ \it $^c$ School of Physics and Optoelectronic Engineering, Beijing University of Technology, Beijing 100124, P.R. China }

\end{center}
\vskip1cm
\begin{abstract}
{\normalsize{}
     We calculate the form factors for the baryon number violation processes of a heavy-flavor baryon decaying into a pseudoscalar meson and a lepton. In the framework of the Standard Model effective field theory, the leptoquark operators at the bottom quark scale, whose matrix elements define the form factors, are derived by integrating out the high energy physics. Under the QCD factorization approach, the form factors of the baryon number violation processes at leading power can be factorized into the convolution of the long-distance hadron wave functions as well as the short-distance hard and jet functions representing the hard scale and hard-collinear scale effects, separately. Based on measurements of the baryon number violation processes by LHCb, we further impose constraints on the new physics constants of leptoquark operators.
 }
\end{abstract}
\end{titlepage}



\section{Introduction}
In the Standard Model, the baryon number ($B$) and the lepton number ($L$) are strictly conserved. 
Sakharov's three conditions \cite{Sakharov:1967dj} state that the C and CP violation, the baryon number violation and the deviation from thermodynamic
equilibrium can explain the matter-antimatter asymmetry in the universe. Therefore, the study of baryon number violation is an essential topic in the search for new physics signals. In new physics beyond the Standard Model, such as the Grand Unified Theory \cite{Pati:1973uk,Georgi:1974sy} and Supersymmetry Theory \cite{Nilles:1983ge},
 there do exist baryon number violation couplings. Since the new physics energy scale is generally far above the top quark mass, we usually utilize the framework of Standard Model effective field theory (SMEFT), in which heavy fields are integrated out to construct the model-independent leptoquark operators \cite{Weinberg:1979sa,Abbott:1980zj,Wilczek:1979hc,Mecaj:2020opd}. 

Experimentally, these baryon number violation couplings are searched through proton decay \cite{Super-Kamiokande:2014otb,JUNO:2022qgr}.
In addition, experimental researchers have also started studying baryon number violation processes in heavy flavor physics at colliders \cite{Zhao:2020bix,BESIII:2019udi,BaBar:2011yks,LHCb:2013fsr,Grunberg:2017key,LHCb:2022wro}, thanks to the improvement of experimental precision. The investigation of leptoquark operators has attracted much attention in heavy flavor physics. In Ref.~\cite{Bauer:2015knc,Ciezarek:2017yzh,Cheung:2020sbq,Huang:2021fuc,Gao:2021sav}, the introduction of leptoquark operators aims to address the anomalies in $R(D)$ and $R(D^{*})$. In the Baryogenesis model \cite{Elor:2018twp,Alonso-Alvarez:2021qfd,Khodjamirian:2022vta,Shi:2023riy}, the form factors of $B$ meson decaying into proton and dark antibaryon have been calculated using light-cone sum rules \cite{Khodjamirian:2022vta}. This calculation seeks to explain baryon number violation and the matter-antimatter asymmetry in the universe. Recently, the BaBar experiment has provided the upper limit for this process \cite{BaBar:2023dtq}.  For processes involving the violation of baryon and lepton numbers, Ref.~\cite{Bansal:2022hhz} calculated the form factors for the decay process $D^{0}\rightarrow \bar{p}e^{+}$ using light-cone sum rules, with the intention of providing theoretical input for the BESIII measurements \cite{BESIII:2021krj}. 

In Ref.~\cite{Grunberg:2017key}, the authors measured the baryon and lepton number violation process 
$\Lambda_{b}\rightarrow K\mu$ in the LHCb experiment and provided an upper limit. To constrain the new physics parameters in conjunction with the experimental data, the theoretical exploration of these decays is indispensable. For the $\Lambda_{b}\rightarrow K\mu$ decay, there exist $\Delta (B-L)=0$ process \cite{Weinberg:1979sa} $\Lambda_{b}\rightarrow K^{-}\ell^{+}$ and $\Delta (B-L)=2$ process \cite{Wilczek:1979et} $\Lambda_{b}\rightarrow K^{+}\ell^{-}$. Our work computes the form factors of spectator processes $\Lambda_{b}\rightarrow P\ell$ for both $\Delta(B-L)=0$ and $\Delta(B-L)=2$, with $P=\pi,\,K$ denoting a light pseudoscalar meson. In the framework of the SMEFT, effective operators of leptoquark at the $m_{b}$ scale are obtained by integrating out the heavy particles at the new physics scale. The matrix elements of the leptoquark operators are calculated in the QCD factorization approach \cite{Beneke:1999br,Beneke:2000ry,Beneke:2000wa,Beneke:2003zv,Grossman:2015cak}, which is widely used in the semi-leptonic and non-leptonic decays of $B$ mesons. At leading power, our calculations indicate that the form factors of the $\Lambda_{b}\rightarrow P\ell$ processes are factorized into the convolution of hard function, jet function, and wave functions without endpoint divergence. The short-distance hard function and jet function, which respectively correspond to the hard scale and hard-collinear scale contributions, could be calculated perturbatively. The form factors of baryon number violation processes can be combined with experimental data to constrain the new physics parameters of leptoquark operators. These form factors can also serve as inputs for calculations in other new physics theories. 

The framework of this paper is as follows: In the next section, we will introduce the effective Hamiltonian and the leptoquark operators. In section \ref{sec:3}, the form factors of $\Lambda_{b}\rightarrow P$ will be calculated in the QCD factorization approach. In section \ref{sec:4}, the numerical result will be given. We present our conclusions in the last section.

\section{leptoquark operator}    \label{sec:2}   

\begin{figure}
	\begin{center}
		\includegraphics[scale=0.35]{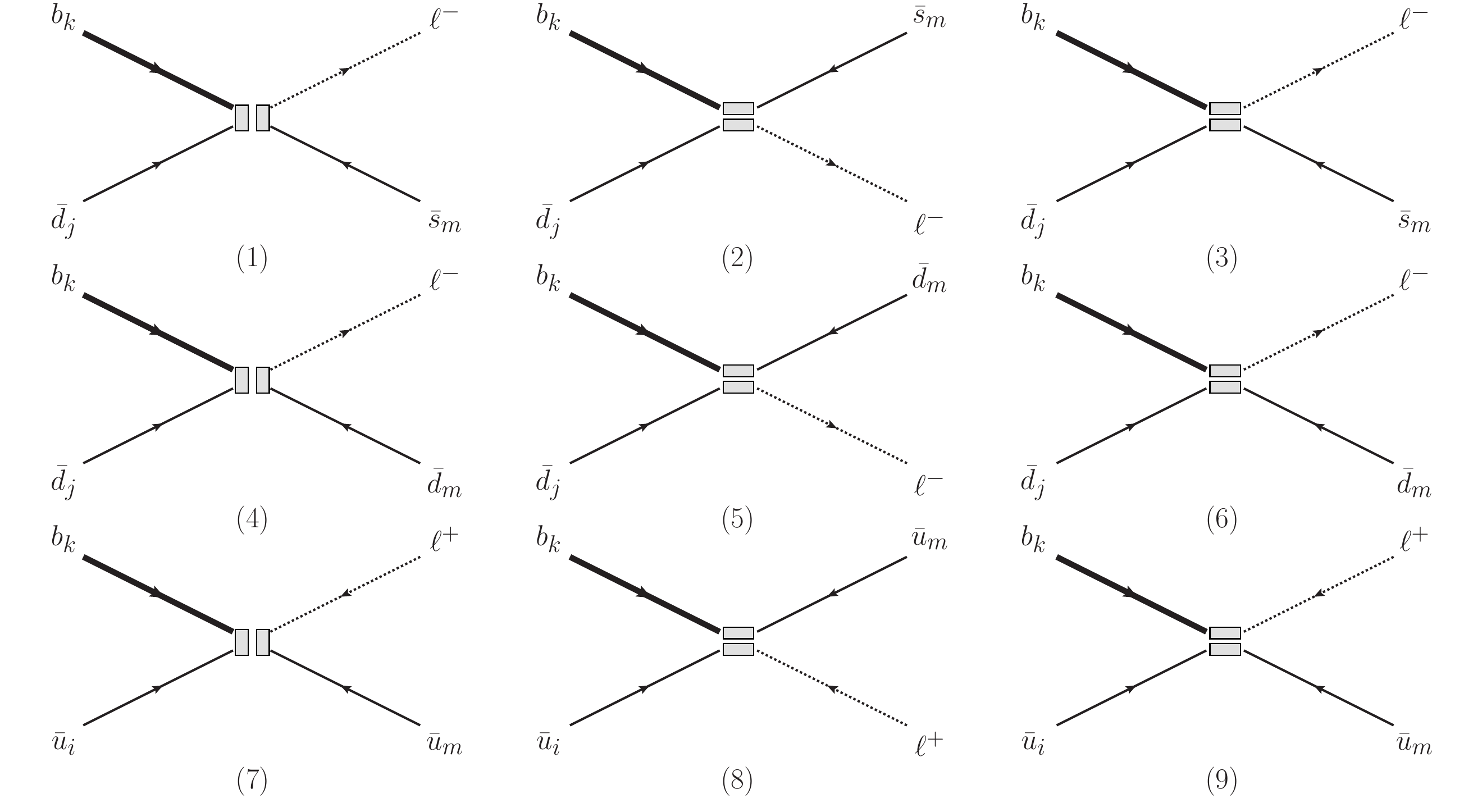}
		\par\end{center}
	\caption{Leptoquark operators of $\Lambda_{b}\rightarrow P \ell$ spectator processes below the $m_{b}$ scale, with the bold lines representing heavy quark fields, the thin lines representing light quark fields, and the dashed lines representing the lepton fields. The shaded blocks in gray color represent the $\Gamma$ matrices connecting two fermion fields. }
	\label{fig:Leptoquark_operator_O}
\end{figure}

In the theory of baryon number violation, the introduction of leptoquark can be traced back to Grand Unification Theories \cite{Pati:1973uk,Georgi:1974sy}, and the leptoquark operator theory \cite{Weinberg:1979sa,Abbott:1980zj,Wilczek:1979hc,Mecaj:2020opd}. In the framework of the SMEFT, the new physics scale is integrated out to obtain the effective Hamiltonian for the $\Lambda_{b}\rightarrow P\ell$ spectator processes as follows
\begin{equation}
\mathcal{H}_{new}=\sum_{\alpha=1}^{7}G_{new,\alpha}O_{\alpha},
\end{equation}
where $G_{new,\alpha}$ represents the effective new physics coupling constants. The corresponding leptoquark operators of spectator processes $\Lambda_b \rightarrow P l$ are
\begin{equation}
\begin{aligned}
	O_{1}  =&\,\epsilon^{ijk}(\bar{d}_{j}^{c}\Gamma b_{i})(\bar{\ell}\Gamma s_{k}),  \qquad 
	O_{2}  =\epsilon^{ijk}(\bar{s}_{i}^{c}\Gamma b_{j})(\bar{\ell}\Gamma d_{k}),&\\
	O_{3}  =&\,\epsilon^{ijk}(\bar{d}_{i}^{c}\Gamma s_{j})(\bar{\ell}\Gamma b_{k}), \qquad
	O_{4}  =\epsilon^{ijk}(\bar{d}_{i}^{c}\Gamma b_{j})(\bar{\ell}\Gamma d_{k}),& \\
	O_{5}  =&\,\epsilon^{ijk}(\bar{d}_{i}^{c}\Gamma d_{j})(\bar{\ell}\Gamma b_{k}),\qquad
	O_{6}  =\epsilon^{ijk}(\bar{u}_{i}^{c}\Gamma b_{j})(\bar{u}^{c}_{k}\Gamma \ell),&\\
	O_{7}  =&\,\epsilon^{ijk}(\bar{u}_{j}^{c}\Gamma u_{i})(\bar{b}^{c}_{k}\Gamma \ell),\qquad
\end{aligned}
    \label{eqn:Leptoquark_operator_O}
\end{equation}
where $\ell=e,\mu$ represent the leptons. $\epsilon^{ijk}$ is a fully antisymmetric tensor and the Latin letters of superscript represent color indices. The violation of fermion flow often occurs in the processes of baryon number violation. Typically, people introduce eigenstates of charge conjugation to construct the leptoquark operators with the specific form
\begin{equation}
	\psi^{c}=C\bar{\psi}^{T},\qquad\bar{\psi}^{c}=\psi^{T}C,
\end{equation}
where $C$ is the matrix of charge conjugation. In Fig.~\ref{fig:Leptoquark_operator_O}, the bold lines represent a heavy quark field, the thin lines represent light quark fields, and the dashed line represents the lepton field. The shaded blocks in gray color represent the $\Gamma$ matrices connecting two fermion fields.  The Lorentz structures of leptoquark operators are represented by
\begin{equation}\label{gammaM}
	\Gamma=\{1,\gamma_{5},\gamma_{\mu},\gamma_{\mu}\gamma_{5},\sigma_{\mu\nu}\},
\end{equation}
where $\sigma_{\mu\nu}=\frac{i}{2}[\gamma_{\mu},\gamma_{\nu}]$. 
For the leptoquark operators $O_4$ and $O_6$, there exist two quarks with the same flavor, which will result in an additional diagram from the Pauli principle. For the convenience of distinguishing different decay channels, we label the colors of $u, d, b$ quark in $\Lambda_{b}$ baryon as $i,j,k$ and $m$ representing the color in the meson to distinguish the different channel.  The diagrams of leptoquark operators are shown in Fig.~\ref{fig:Leptoquark_operator_O}, where diagrams $(1)\sim (3)$ represent operators $O_{1}\sim O_{3}$ respectively. Diagrams $(4)$ and $(5)$ represent the s-channel and t-channel of $O_{4}$ operator, diagram $(6)$ represents the $O_{5}$ operator. Similarly, diagrams $(7)$ and $(8)$ represent the s-channel and t-channel of the $O_{6}$ operator, and diagram $(9)$ represents the $O_{7}$ operator. The leptoquark operators $O_{1}\sim O_{3}$ contribute to the process of $\Lambda_{b}\rightarrow K^{+}\ell^{-}$; $O_{4}$ and $O_{5}$ contribute to the process of $\Lambda_{b}\rightarrow \pi^{+}\ell^{-}$; $O_{6}$ and $O_{7}$ contribute to  the process of $\Lambda_{b}\rightarrow \pi^{-}\ell^{+}$. 
It is easy to see that the operators $O_{1}\sim O_{5}$ violate $B-L$ number, while the operators $O_{6}$ and $ O_{7}$ conserve the $B-L$ number.



\section{QCD calculation}    \label{sec:3}  

In the last section, the leptoquark operators in Eq.~(\ref{eqn:Leptoquark_operator_O})  are obtained at the bottom quark mass scale after integrating out the high-energy physics. 
 The decay matrix elements of $\Lambda_{b}\rightarrow P\ell$ processes are then the product of the  $\Lambda_{b}\rightarrow P$ transition form factor and the lepton spinor. Since the lepton spinor does not couple to the anti-symmetric  $\sigma_{\mu\nu}$, we will not consider the $\Gamma = \sigma_{\mu\nu}$ in Eq.~(\ref{gammaM}). In general,  the $\Lambda_{b}\rightarrow P$ transition form factors can be parametrized as follows 
\begin{equation}
	\begin{aligned}
	\langle P^{+}(p)| O_{\alpha} | \Lambda_{b}(p_{\Lambda_{b}}) \rangle =&\,[\,A^{+}+B^{+}\,\slashed{p}+C^{+}\,\slashed{q}\,]\,u_{\Lambda_{b}}(p_{\Lambda_{b}}),\\
	\langle P^{-}(p) | O_{\alpha} | \Lambda_{b}(p_{\Lambda_{b}}) \rangle =&\,  u^{T}_{\Lambda_{b}}(p_{\Lambda_{b}})\,C\,[\,A^{-}+B^{-}\,\slashed{p}+C^{-}\,\slashed{q}\,],
	\end{aligned}\qquad
\begin{aligned}
    \alpha=1\sim 5,\\
	\alpha=6\sim 7.
\end{aligned}
\end{equation}
According to the equation of motion $\slashed{p}_{\Lambda_{b}}u(p_{\Lambda_{b}})=m_{\Lambda_{b}}u(p_{\Lambda_{b}})$ and the momentum conservation $p_{\Lambda_{b}}=p+q$, where $p$ and $q$ represent the momenta of the final-state meson and lepton, respectively, the scalar form factor $A^{\pm}$ can be decomposed into form factor $B^{\pm}$ and $C^{\pm}$. Therefore, the leptoquark operators have only two independent form factors $B^{\pm}$ and $C^{\pm}$. The form factor $C^{\pm}$ does not contribute to our semi-leptonic decays since the equation of motion $\slashed{q}\,u_{\ell}(q)=0$. 
The only left task for us is the evaluation of the form factor $\zeta_{\Lambda_{b}\rightarrow P^{\pm}}$, which is defined as
\begin{equation}
	\begin{aligned}
		\langle P^{+}(p) \ell^{-}(q) | O_{\alpha} | \Lambda_{b}(p_{\Lambda_{b}}) \rangle \sim &\,\zeta_{\Lambda_{b}\rightarrow P^{+}}\, \bar{u}_{\ell}(q)\,\dfrac{\slashed{n}}{2}\,u_{\Lambda_{b}}(p_{\Lambda_{b}}),\;\;\qquad\alpha=1\sim 5,\\
		\langle P^{-}(p) \ell^{+}(q) | O_{\alpha} | \Lambda_{b}(p_{\Lambda_{b}}) \rangle \sim &\,\,\zeta_{\Lambda_{b}\rightarrow P^{-}}\, u^{T}_{\Lambda_{b}}(p_{\Lambda_{b}})\,C\,\dfrac{\slashed{n}}{2}\,v_{\ell}(q),\,\;\quad\alpha=6\sim 7,
	\end{aligned}
\label{eq:hadron_form_factor}
\end{equation}
where the Lorentz structure between the spinors is from the leading power expansion of the 
meson momentum $\slashed{p}=\bar{n}\cdot p\,\slashed{n}/2$ in light-cone limit, and the definition of the light-cone coordinate system is given in Section \ref{sec:3.1}.

For the processes below the 
$m_{b}$ scale, the QCD factorization approach \cite{Beneke:1999br,Beneke:2000ry,Beneke:2000wa,Beneke:2003zv,Grossman:2015cak} is employed to investigate the hadron decay. The approach provides a factorization scheme for decay amplitudes to be expanded in the power of $\lambda\sim\Lambda_{\mathrm{QCD}}/m_{b}$, which is widely employed in $B$ meson decay. Under the QCD factorization approach, the form factor of heavy-to-light current at leading power can be factorized as follows \cite{Beneke:2000wa}
\begin{equation}
	f_{i}(q^{2})=C_{i}\cdot\xi_{P}(E)+\phi_{B}\otimes T_{i}\otimes\phi_{P}.
	\label{eqn:B_to_V_formfactor}
\end{equation}
In Eq.~(\ref{eqn:B_to_V_formfactor}), the first term on the right-hand side of the equation is the hard function $C_{i}$ times the soft form factor $\xi_{P}(E)$ which absorbs the non-perturbative effects below the hard-collinear scale. Trying to further factorize $\xi_{P}(E)$ into a convolution of a jet function and light-cone distribution amplitudes (LCDAs), one will encounter the well-known endpoint divergence, which destroys the factorization. The second term on the right-hand side of the equation represents the hard-scattering kernel $T_{i}$ convoluted with the LCDAs of the $B$ meson and the light pseudoscalar meson. We will show in the following that for the baryon decay, the form factor could be factorized, thus the factorization formula of the form factor only retains the second term in Eq.~(\ref{eqn:B_to_V_formfactor}) but with the $B$ meson LCDA replaced by the baryon one.

\subsection{Kinematics} \label{sec:3.1}  

Since the $\Lambda_b$ baryon is very heavy, the massless final-state meson and lepton in the baryon number violation processes can be approximated to be on the light-cone. In the light-cone coordinate system, we will introduce two light-cone vectors $n$ and $\bar{n}$
\begin{equation}
	n^{\mu}=(1,0,0,1),\qquad\bar{n}^{\mu}=(1,0,0,-1),
\end{equation}
which satisfy $n^2=\bar{n}^2=0$ and $n\cdot\bar{n}=2$. We choose the large momentum component of the pseudoscalar meson in the collinear direction as 
	$\bar{n}\cdot p=m_{\Lambda_{b}}$ and the large momentum component of the lepton in the anti-collinear direction as $n\cdot q=m_{\Lambda_{b}}$. In the light-cone coordinate, the momentum can be expanded as
\begin{equation}
	k^{\mu}=n\cdot k\dfrac{\bar{n}^{\mu}}{2}+\bar{n}\cdot k\dfrac{n^{\mu}}{2}
	+k_{\perp}^{\mu},
\end{equation}
where we chose the convention $k\sim (n\cdot k,\bar{n}\cdot k,k_{\perp})$. Since the $\Lambda_{b}$ baryon involves a heavy bottom quark, we consider it within the framework of Heavy Quark Effective Theory (HQET) \cite{Manohar:2000dt}, defining the velocity of the heavy baryon
\begin{equation}
	v^{\mu}=(1,0,0,0),
\end{equation} 
satisfying $v^2=1$. For the heavy $b$-quark decay processes in QCD, there exist five different momentum regions
\begin{equation}
\begin{aligned}
		\mathrm{hard}:&\quad(\,1\,,1\,,1\,)\,m_{b},\\
		\mathrm{hard-collinear}:&\quad(\,\lambda,\,1,\,\sqrt{\lambda}\,)\,m_{b},\\
		\mathrm{collinear}:&\quad(\,\lambda^{2},\,1,\,\lambda\,)\,m_{b},\\
		\mathrm{anti-collinear}:&\quad(\,1,\,\lambda^{2},\,\lambda\,)\,m_{b},\\
		\mathrm{soft}:&\quad (\,\lambda,\,\lambda,\,\lambda\,)\,m_{b}.\\
	\end{aligned}
\end{equation}
Except for the heavy quark, the rest of the light degrees of freedom in the baryon have soft momenta. The final-state pseudoscalar meson and lepton represent the collinear mode and anti-collinear mode respectively. The interaction between the collinear field and the soft field is mediated by the hard-collinear field.

\subsection{Light-cone distubation amplitude}  
 To calculate the decay amplitude at leading power, we need to introduce the definition of the LCDAs for the light pseudoscalar mesons at leading-twist \cite{Beneke:2000wa,Grossman:2015cak}
\begin{equation}
	\begin{aligned}
		\left\langle P(p)|\,[\bar{q}(t\bar{n})]_{A}\,[t\bar{n},0]\,[q(0)]_{B}\,|0\right\rangle  & =\dfrac{if_{P}}{4}\,\bar{n}\cdot p\,\bigg[\,\dfrac{\slashed{n}}{2}\,\gamma_{5}\,\bigg]_{BA}\int^{1}_{0}dx\, e^{ixt\bar{n}\cdot p}\,\phi_{P}(x,\mu),\\
	\end{aligned}
\label{eq:pseudoscalar_LCDA}
\end{equation}
where $f_{P}$ is the decay constant of pseudoscalar meson. The leading twist pseudoscalar meson LCDAs can be expanded in terms of Gegenbauer polynomials as
\begin{equation}
	\phi_{P}(x,\mu)=6\,x\,\bar{x}\,\bigg[\,1+\sum_{n=1}^{\infty}a_{n}^{P}(\mu)\,C_{n}^{(3/2)}(2x-1)\,\bigg].
	\label{eq:pseudoscalar_DA_leading_twist}
\end{equation}
where $x$ and $\bar{x}$ are the momenta fraction of the $\bar{q}$ quark and the $q$ quark respectively in the meson, with $\bar{x}=1-x$. The Gegenbauer moments expansion of the pseudoscalar meson is presented up to the first two orders in Table \ref{tab:Input_Para}. The definition of gauge link \cite{Becher:2014oda} is  
\begin{equation}
	[t\bar{n},0]=\mathrm{\bf{P}}\,\mathrm{exp}\bigg[\,ig\int_{0}^{t}dx\,\bar{n}\cdot A(x\bar{n})\,\bigg],
\end{equation}
 connecting along the light-cone direction from $0$ to $t\bar{n}$ at the coordinate space. 
 \begin{table}
	\caption{Imput parameter at $\mu_{0}=1\,\mathrm{GeV}$\label{tab:Input_Para}.}
	\begin{spacing}{1.7}
		\noindent \centering{}%
		\begin{tabular}{ll|ll}
			\hline
			\hline 
			$\quad m_{\Lambda_{b}}=5.6196\,\mathrm{GeV}$ & \cite{ParticleDataGroup:2022pth}	& $\quad\tau_{\Lambda_{b}}=1.471\,\mathrm{ps}$ & \cite{ParticleDataGroup:2022pth} \\
			$\quad\omega_{0}=0.280\,_{-0.038}^{+0.047}\,\mathrm{GeV}$ & \cite{Wang:2015ndk} $\quad$ & 	$\quad f_{\Lambda_{b}}^{(2)}(\mu_{0})=0.030\pm 0.005\,\mathrm{GeV}^{3}$ & \cite{Wang:2015ndk}$\quad$\\
			$\quad f_{\pi}=0.1304\pm0.0002\,\mathrm{GeV}$ & \cite{Grossman:2015cak} $\quad$&	$\quad f_{K}=0.1562\pm0.0007\,\mathrm{GeV}$ & \cite{Grossman:2015cak} \\
			$\quad a_{1}^{\pi}(\mu_{0})=0$ & \cite{Grossman:2015cak} &	$\quad a_{2}^{\pi}(\mu_{0})=0.29\pm0.08$ & \cite{Grossman:2015cak}\\
			$\quad a_{1}^{K}(\mu_{0})=-0.07\pm0.04$ & \cite{Grossman:2015cak} $\quad$ &	$\quad a_{2}^{K}(\mu_{0})=0.24\pm0.08$ & \cite{Grossman:2015cak}\\
			\hline 
			\hline
		\end{tabular}
	\end{spacing}
\end{table}

The definition of the LCDA for the \(\Lambda_{b}\) baryon at leading-twist is as follows \cite{Ball:2008fw,Feldmann:2011xf,Bell:2013tfa,Wang:2015ndk}
\begin{equation}
	\begin{aligned}
		&\left\langle 0|\,[u_{i}(t_{1}n)]_{A}\,[0,t_{1}n]\,[d_{j}(t_{2}n)]_{B}\,[0,t_{2}n]\,[h_{v,k}(0)]_{C}\,|\Lambda_{b}(v)\right\rangle \\ =&\dfrac{\epsilon_{ijk}}{4N_{c}!}\,f_{\Lambda_{b}}^{(2)}(\mu)\,[u_{\Lambda_{b}}(v)]_{C}\,\bigg[\,\dfrac{\slashed{\bar{n}}}{2}\,\gamma_{5}\,C^{T}\,\bigg]_{BA} \int_{0}^{\infty}d\omega\,\omega \int_{0}^{1}dy\, e^{-i\omega(t_{1}y+it_{2}\bar{y})}\,\psi_{2}(y,\omega) ,
	\end{aligned}
\label{eq:Lambda_b_DA}
\end{equation}
where $f^{(2)}_{\Lambda_{b}}$ is the decay constant of $\Lambda_{b}$ baryon. $\omega$ is the sum  of the two light quark $n$-direction momenta. $y$ and $\bar{y}$ correspond to the momentum fraction of $u$ and $d$ quarks respectively, with $\bar{y}=1-y$. The two light quarks   form a di-quark structure \cite{Ball:2008fw} of the  $\Lambda_{b}$ LCDA in Eq.~(\ref{eq:Lambda_b_DA}). $u_{\Lambda_{b}}(v)$ is the Dirac spinor in HQET, satisfying
\begin{equation}
	\slashed{v}\,u_{\Lambda_{b}}(v)=u_{\Lambda_{b}}(v).
\end{equation}  
The definition of gauge link \cite{Wang:2015ndk} is
\begin{equation}
   [0,t_{i}n]=\mathrm{\bf{P}}\,\mathrm{exp}\bigg[\,-ig\int_{0}^{t_{i}}dx\,n\cdot A(xn)\,\bigg],
\end{equation}  
where $i=1,2$. The specific form of the wave function at leading-twist \cite{Feldmann:2011xf,Ball:2008fw,Bell:2013tfa,Wang:2015ndk} is as follows
\begin{equation}
		\psi_{2}(y,\omega) =y\,\bar{y}\,\omega^{2}\,\dfrac{1}{\omega_{0}^{4}}\,e^{-\omega/\omega_{0}},
		\label{eq:Lambda_b_DA_leading_twist}
\end{equation}
where $\omega_{0}$ is a non-perturbative input parameter, whose  numerical value is given in Table \ref{tab:Input_Para}. 

\subsection{$\Lambda_{b} \rightarrow P\ell$ decay amplitude}   

\begin{figure}
	\begin{center}
		\includegraphics[scale=0.303]{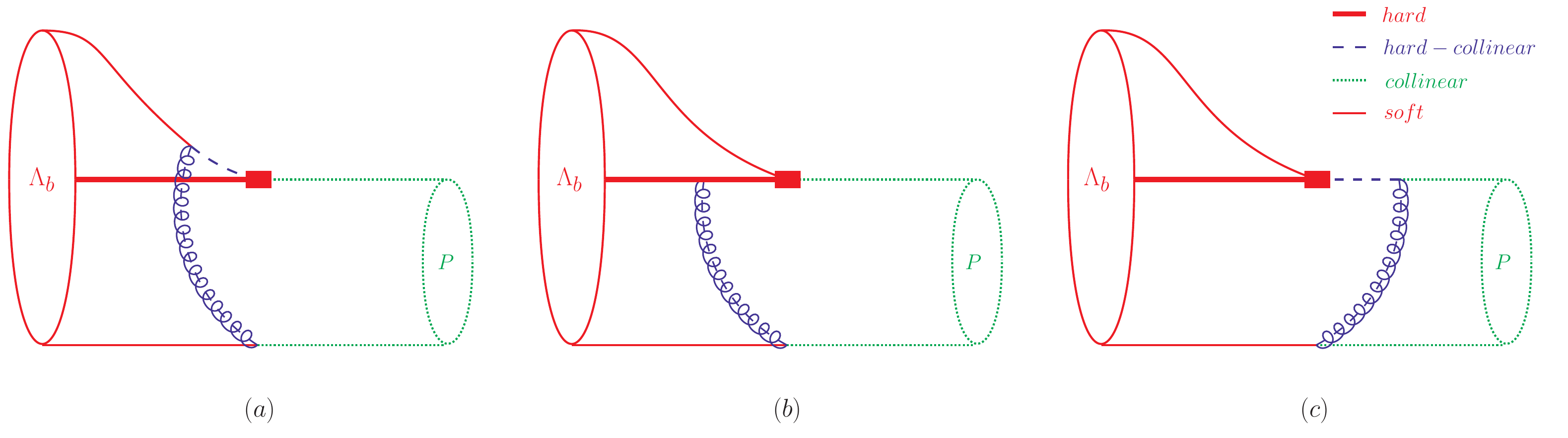}
		\par\end{center}
	\caption{Leading order factorization diagram of $\Lambda_{b} \rightarrow P$ form factor.}
	\label{fig2:QCD_diagrams_total}
\end{figure}

To calculate the form factors of the spectator
processes $\Lambda_{b}\rightarrow P\ell$, we need to construct the correlation functions associated with the leptoquark operators. The leading order Feynman diagrams of the quark field for the baryon number violation processes are shown in Fig.~\ref{fig2:QCD_diagrams_total}. Each interaction vertex in the diagram can be replaced by one of the leptoquark operators from Eq.~(\ref{eqn:Leptoquark_operator_O}). In QCD, the interaction vertex between quark and gluon is 
\begin{equation}
	\mathcal{L}_{int}(x)=\bar{\psi}(x)g\slashed{A}(x)\psi(x).
\end{equation}
Since the operators in Eq.~(\ref{eqn:Leptoquark_operator_O}) involve eigenstates of charge conjugation, we need to introduce an interaction vertex with charge conjugation
\begin{equation}
	\mathcal{L}_{int}^{c}(x)=-\psi^{T}(x)g\slashed{A}^{T}(x)\bar{\psi}^{T}(x).
\end{equation}
According to discrete symmetry, the Lagrangian is invariant under conjugate transformations $\mathcal{L}_{int}(x)=\mathcal{L}_{int}^{c}(x)$. 
The corresponding correlation function in Fig.~\ref{fig2:QCD_diagrams_total} is:
\begin{equation}
	\begin{aligned}
		T_{\alpha} & =\int d^{4}x\,d^{4}y\;T\{O_{\alpha}(0),\,i\mathcal{L}_{int}(x),\,i\mathcal{L}_{int}(y)\}.
	\end{aligned}
    \label{eq:correlation_function}
\end{equation}
In the framework of QCD factorization, we can write down the decay amplitudes for the Fig.~\ref{fig2:QCD_diagrams_total}\,($a$) based on the correlation function:
\begin{equation}
	\begin{aligned}
	   \mathcal{A}_{\alpha}^{a}(\Lambda_{b}\rightarrow P^{+}\ell^{-})&=G_{new,\alpha}\,\zeta^{a}_{\Lambda_{b}\rightarrow P^{+}}\times\bar{u}_{\ell}(q)\,M^{a}_{\alpha}\,u_{\Lambda_{b}}(v),\qquad\,\alpha=1\sim 5,\\
	   \mathcal{A}_{\alpha}^{a}(\Lambda_{b}\rightarrow P^{-}\ell^{+})&=G_{new,\alpha}\,\zeta_{\Lambda_{b}\rightarrow P^{-} }^{a}\times u_{\Lambda_{b}}^{T}(v)\,M_{\alpha}^{a}\,v_{\ell}(q),\qquad\,\alpha=6\sim 7.
	   \label{eqn:Amplitude_Oa}
	\end{aligned}
\end{equation}
where the superscript $a\sim c$ means Fig.~\ref{fig2:QCD_diagrams_total}\,($a$) $\sim$ ($c$) respectively. The form factor $\zeta^{a}_{\Lambda_{b}\rightarrow P^{\pm} }$  at $\mathcal{O}(\alpha_{s})$ can be factored into the decay constants $f_{P}$ and $f_{\Lambda_{b}}$ times the convolution of the hard functions $C^{a}_{\pm}(x,y,\omega,\mu)$, jet functions $\mathcal{J}^{a}_{\pm}(x,y,\omega,\mu)$, and non-perturbative wave functions $\phi_{P}(x,\mu_{0})$ and $\psi_{2}(y,\omega)$ at leading power
\begin{equation}
	\zeta^{a}_{\Lambda_{b}\rightarrow P^{\pm}}=f_{P}f_{\Lambda_{b}}^{(2)}\int_{0}^{1}dx\int_{0}^{\infty}d\omega\,\omega\int_{0}^{1}dy\,C^{a}_{\pm}(x,y,\omega,\mu)\mathcal{J}^{a}_{\pm}(x,y,\omega,\mu)\phi_{P}(x,\mu_{0})\psi_{2}(y,\omega).
	\label{eq:zeta_a}
\end{equation}
The perturbative hard and jet function receives contributions from the hard mode and the hard-collinear mode, respectively. At the tree-level, the hard function $C^{a}_{\pm}(x,y,\omega,\mu)=1$, and the jet functions are
\begin{equation}
	  \mathcal{J}^{a}_{+}(x,y,\omega,\mu)=\dfrac{\pi\,\alpha_{s}(\mu)\, T^{a}_{c}}{4\,\bar{n}\cdot p}\dfrac{1}{x\,y\,\omega^{2}},\qquad
	  \mathcal{J}^{a}_{-}(x,y,\omega,\mu)=\dfrac{\pi\,\alpha_{s}(\mu)\, T^{a}_{c}}{4\,\bar{n}\cdot p}\dfrac{1}{x\,\bar{y}\,\omega^{2}},\\
\label{eqn:jet_function_alpha_a}
\end{equation}
%
where $T^{a}_{c}=2/9$ represents the color factors of leptoquark operators.  The difference between jet functions $\mathcal{J}_{+}^{a}(x,y,\omega,\mu)$ and $\mathcal{J}_{-}^{a}(x,y,\omega,\mu)$ are the momenta fraction $y$ and $\bar{y}$. In the form factors $\zeta^{a}_{\Lambda_{b}\rightarrow P^{\pm}}$, the integral of the jet function $\mathcal{J}_{\pm}^{a}(x,y,\omega,\mu)$ in Eq~(\ref{eqn:jet_function_alpha_a}) and the wave functions $\psi_{2}(y,\omega)$ in Eq.~(\ref{eq:Lambda_b_DA_leading_twist}) with respect to $y$ are  
\begin{equation}
\begin{aligned}
    \int^{1}_{0}dy\,\mathcal{J}_{+}^{a}(x,y,\omega,\mu)\,\psi_{2}(y,\omega)\sim&
    \int^{1}_{0}dy\,\bar{y},\\
    \int^{1}_{0}dy\,\mathcal{J}_{-}^{a}(x,y,\omega,\mu)\,\psi_{2}(y,\omega)\sim&
    \int^{1}_{0}dy\,y,\\
    \end{aligned}
\end{equation}
The momenta fraction $y$ of $u$-quark and $\bar{y}$ of $d$-quark are symmetric under the convolution of $\int_{0}^{1}dy\,y =\int_{0}^{1}dy\,\bar{y}$, since the light quarks of $\Lambda_{b}$ LCDA in Eq.~(\ref{eq:Lambda_b_DA}) are symmetric in heavy quark limit \cite{Bell:2013tfa}. Therefore,
the form factors can be abbreviated as one $\zeta^{a}_{\Lambda_{b}\rightarrow P}=\zeta^{a}_{\Lambda_{b}\rightarrow P^{\pm}}$.

For the decay amplitudes of the baryon number violation processes, the spinor structures are formed by the lepton spinor and the heavy baryon spinor. The difference of the spinor structures between the processes $\Lambda_{b}\rightarrow P^+\ell^-$ and $\Lambda_{b}\rightarrow P^-\ell^+$ arises from whether the fermion flow of the leptoquark operators in Eq.~(\ref{eqn:Leptoquark_operator_O}) is entirely broken. The fermion flows in the lepton part of operators $O_{1}\sim O_{5}$ are not broken, hence the spin structure is the bilinear form $\bar{u}_{\ell}(q)\,M_{\alpha}^{a}\,u_{\Lambda_{b}}(v)$. On the other hand, the fermion flows in the lepton part of the operators $O_{6}$ and $ O_{7}$ are broken, resulting in a spin structure in the form of $u_{\Lambda_{b}}^{T}(v)\,M_{\alpha}^{a}\,v_{\ell}(q)$. $M^{a}_{\alpha}$ represents the matrix element between the heavy baryon spinor field  and the lepton spinor field 
\begin{equation}
\begin{aligned}
	M_{1}^{a} & =-M_{2}^{a}=-M_{4}^{a}/2=-2\,\bar{n}\cdot p\,m_{\Lambda_{b}}^{1/2}\,\Gamma\,\dfrac{\slashed{n}}{2}\,\Gamma,\\
	M_{6}^{a} & =4\,\bar{n}\cdot p\,m_{\Lambda_{b}}^{1/2}\,\Gamma^{T}\,C\,\dfrac{\slashed{n}}{2}\,\Gamma,\\
\end{aligned}\qquad
\begin{aligned}
	M_{3}^{a} & =M_{5}^{a}=2\,\bar{n}\cdot p\,m_{\Lambda_{b}}^{1/2}\,\mathrm{Tr}\bigg\{\Gamma\,\dfrac{\slashed{n}}{2}\bigg\}\,\Gamma,\\
	M_{7}^{a} & =2\,\bar{n}\cdot p\,m_{\Lambda_{b}}^{1/2}\,\mathrm{Tr}\bigg\{\Gamma\,\dfrac{\slashed{n}}{2}\bigg\}\, C\,\Gamma,
\end{aligned}
\label{eq:Ma}
\end{equation}
where $m_{\Lambda_{b}}^{1/2}$ is from the difference of the spinor between QCD and HQET $u_{\Lambda_{b}}(p_{\Lambda_{b}}) = m_{\Lambda_{b}}^{1/2} u_{\Lambda_{b}}(v)$. 
From Eq.~(\ref{eqn:Leptoquark_operator_O}) and Fig.~\ref{fig:Leptoquark_operator_O}, one can see that for operators $O_3,O_5$ and $O_7$, the Dirac fermion flow is a $b$-quark going to a lepton, while the light quark in these operators combines with spectator quark going to the final-state pseudoscalar meson via the strong and the new physics interaction. The pseudoscalar mesons are formed by the di-quark structure, which is similar to the semi-leptonic decays $B\rightarrow\pi \ell\nu$  \cite{Beneke:2000wa} or radiative decays $B\rightarrow V\gamma $  \cite{Becher:2005fg,Deng:2021zoi}. In this case, the calculation will give a contribution as a trace of the Dirac matrices shown in the second column of Eq.~(\ref{eq:Ma}). Because of this trace in Eq.~(\ref{eq:Ma}), only $\Gamma=\gamma_\mu$ kind of operators can contribute to our calculation, which is similar to the $B$ decay case \cite{Beneke:2000wa,Bauer:2002aj,Becher:2005fg,Deng:2021zoi}.
For the other kinds of operators, shown in the first column of Eq.~(\ref{eq:Ma}), no trace is required, such that all kinds of operators contribute except  that for  $\Gamma =\sigma_{\mu\nu}$, since the lepton spinor does not couple to the anti-symmetric  $\sigma_{\mu\nu}$ at leading power. 
 
Kinematically, the above calculation can be described by the factorization diagrams for the form factors of the baryon number violation processes, as depicted in Fig.~\ref{fig2:QCD_diagrams_total}\,($a$). The $b$-quark decays through new physics particle into two antiquark fields and a lepton field, with one of the hard-collinear antiquark fields annihilating with the soft quark from the initial state to produce a hard-collinear gluon, which converts the soft spectator quark into a collinear one. The jet function in Eq.~(\ref{eqn:jet_function_alpha_a}) is obtained when the hard-collinear fields are integrated out, which gives a leading power contribution. 

Comparing  Fig.~\ref{fig2:QCD_diagrams_total}\,($a$) and ($b$), the light quark propagator is replaced by a heavy quark propagator in the perturbative calculation. Thus the contribution from Fig.~\ref{fig2:QCD_diagrams_total}\,($b$) is suppressed by $\Lambda_{\mathrm{QCD}}/m_b$  compared with Fig.~\ref{fig2:QCD_diagrams_total}\,($a$). We will not consider this next-to-leading power contribution in the present paper. 

Based on the correlation function in Eq.~(\ref{eq:correlation_function}), we can similarly express the amplitudes of Fig.~\ref{fig2:QCD_diagrams_total}\,($c$) as
\begin{equation}
	\begin{aligned}
	\mathcal{A}_{\alpha}^{c}(\Lambda_{b}\rightarrow P^{+}\ell^{-})&=G_{new,\alpha}\,\zeta^{c}_{\Lambda_{b}\rightarrow P^{+} ,\alpha}\times\bar{u}_{\ell}(q)\,M^{c}_{\alpha}\,u_{\Lambda_{b}}(v),\qquad\alpha=1\sim 5,\\
	\mathcal{A}_{\alpha}^{c}(\Lambda_{b}\rightarrow P^{-}\ell^{+})&=G_{new,\alpha}\,\zeta_{\Lambda_{b}\rightarrow P^{-} ,\alpha}^{c}\times u_{\Lambda_{b}}^{T}(v)\,M_{\alpha}^{c}\,v_{\ell}(q),\qquad\alpha=6\sim 7.
	\end{aligned}
	\label{eqn:factorization3}
\end{equation}
It is easy to show that at leading power, the matrix element $M^{c}_{\alpha}$ between spinors for any of the effective operators for this Feynman diagram are proportional to 
\begin{equation}
		M_{\alpha}^{c}    \sim  \dfrac{\slashed{n}}{2}\,\gamma^{\mu}\,\dfrac{\slashed{n}}{2}\,\gamma_{\mu}\,\dfrac{\slashed{\bar{n}}}{2} = \dfrac{\slashed{\bar{n}}}{2}\,\gamma^{\mu}\,\dfrac{\slashed{n}}{2}\,\gamma_{\mu}\,\dfrac{\slashed{n}}{2} = 0.
\end{equation}
In fact, this diagram is similar to the one in semi-leptonic $B$ decay, which has the same power as Fig.~\ref{fig2:QCD_diagrams_total}\,($b$). Remarkably, the amplitudes of Fig.~\ref{fig2:QCD_diagrams_total}\,($b$) and ($c$) are  the major contribution of the $B$ meson heavy-to-light form factors at leading power \cite{Beneke:2000wa,Bauer:2002aj,Deng:2021zoi}, which are power suppressed in our baryon decay processes.  As a result, only Fig.~\ref{fig2:QCD_diagrams_total}\,($a$) will contribute to the decay amplitudes at leading power and the form factors in Eq.~(\ref{eq:hadron_form_factor}) are $\zeta_{\Lambda_{b}\rightarrow P}=\zeta^{a}_{\Lambda_{b}\rightarrow P}+\mathcal{O}(\lambda)$.
The baryon decay amplitudes are free of endpoint divergence since the convolution of the jet function with the LCDAs in Eq.~(\ref{eq:zeta_a}) is convergent.
At the endpoint region, the LCDAs display the following asymptotic behaviors $\phi_P\mathop{\sim}\limits^{x\to 0} x$, $\psi_2 \mathop{\sim}\limits^{y\to 0,\omega\to 0} y\omega^2$ and $\psi_2 \mathop{\sim}\limits^{y\to 1,\omega\to 0} \bar y\omega^2$ which will compensate the endpoint-divergent behavior $\mathcal{J}^{a}_{+}(x,y,\omega,\mu)\sim 1/(xy\omega^2)$ and $\mathcal{J}^{a}_{-}(x,y,\omega,\mu)\sim 1/(x\bar{y}\omega^2)$ carried by the jet function, and thus the form factor in Eq.~(\ref{eq:zeta_a}) is endpoint finite.
This conclusion is in agreement with the heavy baryon transition form factors in the standard model case \cite{Wang:2011uv}.

  With the transition form factors in Eq.~(\ref{eq:zeta_a}), we can get the decay rate   of $\Lambda_{b}\rightarrow P\ell$ after the sum of spins of final states and averaging over the spin of $\Lambda_b$ for each effective operator. Since the form factors defined for vector current in Eq.(\ref{eq:hadron_form_factor}), we can get the decay rate for  $\Gamma=\gamma_{\mu}$ of each leptoquark operators  in Eq.~(\ref{eqn:Leptoquark_operator_O}) as
 \begin{equation}
\begin{aligned}
\Gamma_{\alpha}^{\mathrm{V}} &=\dfrac{m_{\Lambda_{b}}^{3}}{\pi}\,|G_{new,\alpha}\,\zeta_{\Lambda_{b}\rightarrow K}|^{2},\;\;\qquad\alpha=1\sim2,\\
\Gamma_{\alpha}^{\mathrm{V}} &=\dfrac{4\,m_{\Lambda_{b}}^{3}}{\pi}\,|G_{new,\alpha}\,\zeta_{\Lambda_{b}\rightarrow K}|^{2},\qquad\alpha=3,\\
\Gamma_{\alpha}^{\mathrm{V}} & =\dfrac{4\,m_{\Lambda_{b}}^{3}}{\pi}\,|G_{new,\alpha}\,\zeta_{\Lambda_{b}\rightarrow\pi}|^{2},\qquad\;\alpha=4\sim7,
\end{aligned}
\label{eq:deacy_width_V}
\end{equation}
with superscript V denoting $\Gamma=\gamma_{\mu}$.
For effective operators $O_3$, $O_5$ and $O_7$, the trace term $\mathrm{Tr}
\{\Gamma\,\slashed{n}/2$\} in Eq.~(\ref{eq:Ma}) is equal to zero when $\Gamma=1,\gamma_{5},\gamma_{\mu}\gamma_{5}$, which means that these Lorentz structures will not contribute at leading power for these three operators. 
For other operators, the  $\Gamma=1,\gamma_{5},\gamma_{\mu}\gamma_{5}$ of leptoquark operators $O_{\alpha}$ will also contribute, thus the decay rates labeled S,\,P,\,A in superscript to denote these gamma matrix, are  
\begin{equation}
\begin{aligned}\Gamma_{\alpha}^{\mathrm{S}}=\Gamma_{\alpha}^{\mathrm{P}}=\dfrac{1}{4}\,\Gamma_{\alpha}^{\mathrm{A}} & =\dfrac{m_{\Lambda_{b}}^{3}}{4\text{\,}\pi}\,|G_{new,\alpha}\,\zeta_{\Lambda_{b}\rightarrow K}|^{2},\qquad\alpha=1\sim2,\\
\Gamma_{\alpha}^{\mathrm{S}}=\Gamma_{\alpha}^{\mathrm{P}}=\dfrac{1}{4}\,\Gamma_{\alpha}^{\mathrm{A}} & =\dfrac{m_{\Lambda_{b}}^{3}}{\pi}\,|G_{new,\alpha}\,\zeta_{\Lambda_{b}\rightarrow\pi}|^{2},\qquad\;\alpha=4,\; 6.
\end{aligned}
\label{eq:deacy_width_SPA}
\end{equation}

\section{Numerical result}  \label{sec:4}

\begin{figure}
	\begin{center}
		\includegraphics[scale=0.9]{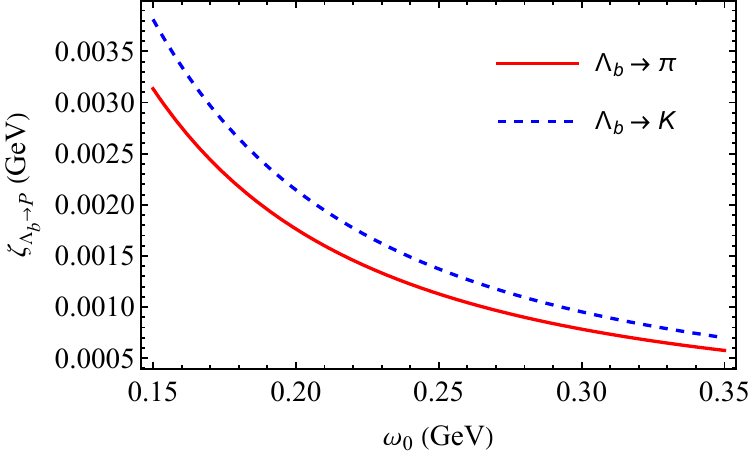}
		\par\end{center}
	\caption{The input parameter $\omega_{0}$ dependence of $\Lambda_{b} \rightarrow P$ form factor.}
	\label{fig:zeta_pi_K}
\end{figure}

In the previous section, we provided the factorized formula of the form factors for the baryon number violation processes. 
Taking into account the input parameters from Table \ref{tab:Input_Para}, we can choose the strong coupling constant at hard-collinear scale  $\alpha_{s}(\mu=\mathrm{2\,GeV})\simeq 0.3$ and calculate the form factors for the spectator processes $\Lambda_{b}\rightarrow P^{+}\ell^{-}$ and $\Lambda_{b}\rightarrow P^{-}\ell^{+}$. The numerical results of $\zeta_{\Lambda_{b}\rightarrow K}$ and $\zeta_{\Lambda_{b}\rightarrow \pi}$ form factors are
\begin{equation}
\begin{aligned}
    \zeta_{\Lambda_{b}\rightarrow K}=&\,1.09\,^{+0.36}_{-0.42}\times 10^{-3}\, \mathrm{GeV},\\
    \zeta_{\Lambda_{b}\rightarrow\pi}=&\,9.00\,^{+2.92}_{-3.42}\times 10^{-4}\, \mathrm{GeV}.
\end{aligned}
    \label{eq.28}
\end{equation}

It is easy to see that there is only one independent form factor at the heavy quark limit for each kind of decay, just like the $B$ meson decays \cite{Beneke:2000wa,Bauer:2002aj,Becher:2005fg,Deng:2021zoi}.
During the form factor calculation, the primary source of theoretical uncertainties in Eq.~(\ref{eq.28}) originates from the non-perturbative input parameter $\omega_{0}$ of the $\Lambda_{b}$ wave function. The dependence of form factors  $\zeta_{\Lambda_{b}\rightarrow K}$ and $\zeta_{\Lambda_{b}\rightarrow \pi}$ on this non-perturbative input are shown in Fig.~\ref{fig:zeta_pi_K}. It can be observed that the influence of the non-perturbative input $\omega_{0}$ on the form factor $\zeta_{\Lambda_{b}\rightarrow P}$ is monotonically decreasing. 
In addition to the above theoretical uncertainty from non-perturbative parameters, the  next-to-leading order QCD correction to the  form factor is at the order of $\alpha_s/\pi \sim  10\%$ and the power correction is estimated at the order of $1/m_b \sim 20\%$.

The LHCb experiments have ever searched for the processes $\Lambda_{b}\rightarrow K^{+}\mu^{-}$ and $\Lambda_{b}\rightarrow K^{-}\mu^{+}$  \cite{Grunberg:2017key}, with the joint upper limit on the branching ratios of 
\begin{equation}
	[\mathcal{B}(\Lambda_{b}\rightarrow K^{-}\mu^{+})+\mathcal{B}(\Lambda_{b}\rightarrow K^{+}\mu^{-})]\times\dfrac{3.1\times10^{-6}}{\mathcal{B}(\Lambda_{b}\rightarrow pK^{-})}<1.95\times10^{-9}\;\mathrm{at}\;\mathrm{CL}=90\,\%.
\end{equation}
Combining the branching ratio  $\mathcal{B}(\Lambda_{b}\rightarrow pK^{-})=(5.4\pm1.0)\times 10^{-6}$ provided by the particle data group \cite{ParticleDataGroup:2022pth}, we can obtain the constraints of new physics couplings for leptoquark operators. 
As stated in section 2, the leptoquark operator $O_{1}\sim O_{3}$ contribute to the process of $\Lambda_{b}\rightarrow K^{+}\ell^{-}$.
Assuming that only one kind of operator contributes to the corresponding decay channel, constraints on new physics couplings $G_{new,\alpha}$ for different Lorentz structures are presented in Table \ref{tab:Branch-ratio-of-non-local}.
As discussed in the previous section, only vector current contributes to the effective operators $O_3$, $O_5$ and $O_7$, so the constraint to operator $O_3$ only occurs for the vector current case.  The tensor structure does not contribute to any kind of effective operators, so there is no constraint from the current experiment. As for operators $O_1$ and $O_2$, we have two kinds of constraints: one is for 
$1$ or $\gamma_{5}$ operators, the other is for $\gamma_{\mu}$ or $\gamma_{\mu}\gamma_{5}$ currents.

\begin{table}
	\caption{Constraints of the new physics effective couplings from LHCb experiments. \label{tab:Branch-ratio-of-non-local}}
	\begin{spacing}{1.7}
		\noindent \centering{}%
		\begin{tabular}{ccccc}
			\hline
			\hline 
			\textbf{$|G_{new,\alpha}|^{2}\; [\mathrm{GeV^{-4}}]$} & \textbf{$\Gamma=1$} or \textbf{$\gamma_{5}$}& \textbf{$\Gamma=\gamma_{\mu}$} & \textbf{$\Gamma=\gamma_{\mu}\gamma_{5}$}  & \textbf{$\Gamma=\sigma_{\mu\nu}$} \tabularnewline
			\hline 
			$\alpha=1,2$ &		$<5.2\times10^{-17}$ &	$<1.3\times10^{-17}$ &	$<1.3\times10^{-17}$	&-\\
			$\alpha=3$&  - & $<3.2\times10^{-18}$ & - & - \\
			\hline 
			\hline
		\end{tabular}
	\end{spacing}
\end{table}

\section{Conclusion} \label{sec:5}

In this work, we have introduced the factorization theorems for baryon and lepton number violation processes in heavy baryon decay. Within the framework of the SMEFT, we integrated out the new physics particles and absorbed them into effective new physics couplings $G_{new,\alpha}$, obtaining leptoquark operators at the $m_{b}$ scale. In the QCD factorization approach, we factorized the form factors for the processes $\Lambda_{b}\rightarrow P\ell$ into the convolution of the hard function, jet function, and wave functions. The combination of the effective field theory of the Standard Model and QCD factorization allows us to understand the factorization behavior of baryon number violation processes even in a low-energy situation. We computed the numerical results of the form factors for these effective operators, which can also be applied to similar baryon number violation processes or serve as inputs for other leptoquark new physics theories. 

Finally, as an example of application, utilizing measurements of $\Lambda_b\to K^\pm \mu^\mp$ decays from LHCb experiments, we derive constraints on some of the effective new physics couplings of the leptoquark operators. With the upgrade and renovation of the High-Luminosity Large Hadron Collider, future experiments will be able to conduct more precise measurements of baryon number violation processes.

\section*{Acknowledgements}

We are grateful to Dong-Hao Li for helpful discussions. The work is partly supported by the National Natural Science Foundation of China with Grant No.12275277 and the National Key Research and Development Program of China under Contract No.2020YFA0406400 and 2023YFA1606000.

\bibliographystyle{apsrev4-2}
\addcontentsline{toc}{section}{\refname}\bibliography{
reference}

	\end{sloppypar}
\end{document}